\documentstyle[aps,prd]{revtex} 
\draft

\begin{document}
\twocolumn[\hsize\textwidth\columnwidth\hsize\csname
@twocolumnfalse\endcsname
\title{%
\hbox to\hsize{\normalsize\rm February 1998, revised April 1998
\hfil Preprint MPI-PTh/98-18}
\vskip 36pt
When The  Wavepacket Is Unnecessary}
\author{L.~Stodolsky}
\address{Max-Planck-Institut f\"ur Physik 
(Werner-Heisenberg-Institut),
F\"ohringer Ring 6, 80805 M\"unchen, Germany}
\date{February 20, 1998}
\maketitle
\begin{abstract}
  We point out that many wavepacket discussions for the coherence
  properties of particle beams are unnecessary since they deal with
  stationary sources; and when the problem is stationary, essentially
  all information is in the energy spectrum. This recognition allows a
  simple answer to a number of long-debated points, usually framed in
  terms of ``length of the wavepacket.'' In particular we discuss
  neutrino oscillations, and some issues in neutron physics. The
  question as to whether two simple beams with the same energy
  spectrum are distinguishable is answered negatively for stationary
  situations.  The question as to whether neutrino oscillations should
  be thought of as taking place between states of the same energy or
  the same momentum is answered in favor of energy for stationary
  situations.  Consequences for proposals involving the $^7$Be
  neutrino line of the sun, the observation of oscillations in
  supernova neutrinos and wavepacket studies with neutrons are briefly
  discussed, as well as the connection with the coherence notions of
  quantum optics.
\end{abstract}
\vskip2.0pc]


\section{Introduction}

A number of apparently subtle and difficult issues, often involving
the concept of ``length of the wavepacket,'' have long been discussed
concerning the coherence properties of various particle beams.

In connection with the possibility of neutrino interference and
oscillations, for example, such issues are often discussed, and there
are many papers and books~\cite{pevsner} where it is treated.  In
particular there has been an extensive discussion around the
suggestion~\cite{pontec} that neutrino oscillation effects might
appear in the annual variation of the earth-sun distance.  For such an
effect to occur, firstly, the neutrino mixing parameters must be in a
favorable range.  Secondly, by using an essentially monoenergetic
source, the electron-capture $^7$Be neutrino ``line'' from the sun, it
is hoped that a washing out of the sought-for oscillations due to
their energy dependence could be avoided. However, there are various
line broadening and possibly other effects, and it seems the coherence
properties of the neutrino flux must be understood. These have been
examined in terms of the ``length of the wavepacket'' resulting from
the electron capture process, first by Nussinov~\cite{nuss} and more
recently re-examined~\cite{nussa} by him and collaborators. Other
authors~\cite{wilz} have also looked at the point in the same way but
have disagreed with some of the conclusions.

In a similar vein, we~\cite{reinartz} tried to assess the
observability of oscillation effects for the neutrinos from a
supernova. This possibility arises if one envisions very small
neutrino mass differences. ``Normally,'' that is when the mass
differences involved are not very small, one would expect the
different mass eigenstates to separate into distinct pulses, due to
their differing velocities and the great distance to the earth.
Nevertheless, for extremely small mass differences the pulses could
overlap upon arrival, suggesting possible oscillation effects.
However, it was difficult to pursue the matter since we were uncertain
what ``length of the wavepacket'' to use.

Again, in neutron physics, where the coherence properties of particle
beams can be particularly well studied, there have been several
discussions as to whether and how it could be possible to determine or
observe the wavepacket properties of a beam~\cite{flow,golub}.

In this note we would like to point out that in such problems the
concern about wavepackets was actually unnecessary. For a {\it
stationary\/} system all information necessary for single particle
measurements is contained in the energy spectrum; and the sun or
a reactor or even a supernova for most purposes,  can certainly be
regarded as essentially stationary~sources.

This recognition also allows us to address a question which arises in
these discussions: Is it possible to tell the difference between a
simple beam consisting of a mixture of long wavepackets, where one
might suppose a high degree of spatial coherence and an apparently
more incoherent beam made out of a mixture of short wavepackets, given
that both beams have the same energy spectrum?  (By ``simple'' we mean
there is no non-trivial subspace, as for mixing, see below.)  This
question was discussed in Refs.~\cite{flow}, and \cite{golub} in
connection with neutron physics, and in \cite{nussa} in connection
with neutrinos. On the basis of various arguments and examples it was
concluded that such a distinction is not possible, at least in
practical experiments, although in Ref.~\cite{nussa} the theoretical
possibility was left open.

Our conclusion will be that such a distinction is never possible, in
single particle quantum mechanics and for stationary situations.  This
is because there is in reality no difference between the two beams.

\section{Stationarity and Energy Spectrum}
 
We shall describe the beam of particles in question by means of a
density matrix $\rho$. We make two important assumptions. First we
assume that we may use a single particle description; $\rho$ is the
density matrix for a single particle. This assumption is not
essential, we make it in order to simplify our discussion of
stationarity and to stay in the same language as that of the
discussions refered to above in neutrino and neutron physics.  Thus
for the moment we consider only experiments involving single particle
counting, and ignore questions connected with correlations between
counts, statistics effects, and multiply occupied states.  However,
precisely these questions are important in quantum optics, and below
we examine the relation to it. Note that for real neutrino or neutron
beams the density is always so low that such effects are unimportant
in any case.

Secondly, and more centrally, we assume that the problem and thus the
density matrix is stationary. By this we mean that no measurement on
the beam described by this density matrix can show a time dependence,
unless of course there are time dependent elements in the measuring
device itself.

To see the implications of this assumption, we recall that the density
matrix is written in terms of single particle wavefunctions $\psi$ for
the beam particle as $\rho= \sum w_i \psi_i\psi_i^*$, where $w_i$ is
the ``weight'' for a state $i$. Let us consider the time dependence of
$\rho$. The weights $w$ represent the properties of the source; they
will be constant if we assume the source is stationary, which we do.
This leaves the time development of the $\psi_i$ as the origin of a
possible time dependence, as given by the usual~equation
\begin{equation}\label{one}
\dot\rho=-i[H,\rho]\,,
\end{equation}
where $H$ is the free Hamiltonian for the  beam particle. 

Now, if $\rho$ must be constant the above equation says that $H$ and
$\rho$ commute. If $\rho$ were not constant, then there would be in
principle some measurement which would show a time variation.  It
seems self-evident for thermal sources like the sun or sources like
reactors that no such measurement is possible. (However the notion of
stationarity may not be trivial when we go beyond the single particle
problem; see the remarks below on the coherent or Glauber state.)

Now since $\rho$ commutes with $H$, it can be diagonalized in the
energy basis. This means that up to obvious degeneracies such as
direction or polarization, the beam is entirely characterized by the
diagonal elements of $\rho$ in the energy basis. But this is simply
the energy spectrum. In other words, $\rho$ is entirely determined by
the energy spectrum.

We conclude that given an energy spectrum and stationary conditions,
no detailed discussion of production mechanisms is necessary.
Furthermore, two stationary beams with the same energy spectrum have
the same density matrix and so cannot be distinguished (by single
counting experiments, see below).

\section{ Time Averaging by the Detector}
Even if the beam is time dependent, there will be many cases where
this time dependence plays no role. Experiments with a chopper in a
neutron beam or at a pulsed accelerator are obviously equivalent to
those with a continuous beam, if time-of-flight or other timing
information is not used. If the detector contains no time-dependent
elements, that is performs a time average, any effects involving
off-diagonal elements of $\rho$, $\rho_{E,E'}\sim e^{i(E-E')t}$ will
be averaged to zero by the integration over time. Then, as for the
stationary beam, only the diagonal elements of $\rho$ enter into the
result and all relevant information is again given by the energy
spectrum.

On the other hand, if the detector has time-dependent elements, as
when we use timing information, these may ``beat" with $e^{i(E-E')t}$
so that off-diagonal elements indeed play a role. Implicit in these
arguments is the assumption that any output or result is linear in the
input density matrix, but this is a fundamental feature of quantum
mechanics.

\section{Density Matrix}

The bothersome feeling that more than just the spectrum ought to be
involved is perhaps traceable to the somewhat unintuitive character of
the density matrix in quantum mechanics. A single, given, density
matrix can arise in different ways, especially when incoherence is
involved. An unpolarized spin 1/2 object is equally well a mixture of
spin-up and spin-down states on the one hand or a mixture of spin-left
and spin-right on the other; an unpolarized photon beam is just as
well a mixture of two linear or a mixture of two circular
polarizations; and so on.

Similarly in the present problem: the main point is the absence of
off-diagonal energy correlations in the density matrix. This might be
thought of as arising in various ways; nevertheless once we know that
the density matrix is stationary the results of these different ways
are all equivalent.  That is, given a stationary density matrix $\rho=
\sum w_i \psi_i\psi^*_i$ and a stationary density matrix $\rho'$ made
up of different states and different weights $\rho'= \sum w'_i
\psi'_i\psi'^*_i$ but in such a way that both have the same energy
spectrum, the two are in fact equal, $\rho=\rho'$.  Thus there is no
way---at least in the usual understanding of quantum mechanics---to
ascribe a difference between a stationary, single particle beam which
is a mixture of short wavepackets on the one hand and one which is a
mixture of near-plane waves on the other, if the energy spectra are
the same.

Since the same density matrix may be thought of as arising in various
ways, for practical considerations, we can view it in the most
convenient form. This will usually be as a mixture of energy
eigenstates for stationary problems, and in the following we always
take an incoherent average over the energy spectrum.

\section{Particle Mixing}

We now turn to particle mixing, as for neutrinos, $K^0$'s or other
neutral heavy flavor mesons.

{\it Energy or momentum?}---In mixing problems, where we have to deal
with linear combinations of particles of different mass, the question
comes up as to whether one should deal with states of the same energy
or the same momentum. Since as stated above, for stationary conditions
we are to perform the calculation as an incoherent sum over energies,
we have given the answer ``energy.''  Evidently, for stationary
problems it is most natural to use stationary wavefunctions $\sim
e^{-iEt}$.

{\it Non-Trivial Subspace.}---But here the density matrix, although
diagonal in energy for stationary conditions, will have a non-trivial
subspace for a given energy. That is, the density matrix element at a
given energy will in general be a matrix, say a $2\times2$ matrix for
a two state system, and in many cases this matrix will be non-trivial.
Thus for mixing problems we must qualify our statement that ``all
information is in the energy spectrum'' and consider how to determine
this matrix.

The most commonly discussed situation is that of emission of a state
with a definite quantum number (``flavor'') but which does not
necessarily correspond to a definite mass. We concentrate on this
case. In contrast to the kinematic variables, we then have a pure
state with respect to the internal (``flavor'') variables. We may
have, for instance, that the neutrino is emitted by nuclear
beta-decay, as a $\nu_e$; or for kaons with some flavor tag, as say a
$\bar K^0$. The problem now is to determine this pure state. We do
this by using that fact it is fixed at emission~\cite{lipkin}, at
$z=0$, where $z$ is the spatial coordinate.  For a two-state system as
with two neutrino species or with $K^0$'s, we have in terms of
ordinary spatial wavefunctions $e^{ip z}$ and internal spinors U:
\begin{equation}\label{threeB}
\psi= \alpha U_1 e^{ip_1 z} +\beta U_2e^{ip_2 z}
\end{equation}
where 1 and 2 refer to the mass eigenstates so $p_1^2=E^2-m_1^2$ and
$p_2^2=E^2-m_2^2$, and the $U$'s are the mass eigenstates.  The
coefficients $\alpha$ and $\beta$ are now so chosen that $\alpha U_1
+\beta U_2$ give the desired state at emission, and so oscillations
take place in space due to the non-vanishing difference between $p_1$
and $p_2$.  Oscillation effects at a given detection point are then
calculated as an incoherent average over $E$.

{\it Decoherence.}---In the foregoing case we had a pure state for
each energy, given by the wavefunction Eq.~(\ref{threeB}) and so a
correspondingly simple density matrix in the $2\times2$ subspace.  Now
the density matrix should in principle be calculated from the details
of the various production processes.  One may ask the following
question ~\cite{hjl}: Suppose it were possible to distinguish which
neutrino mass eigenstate is emitted by detection of the recoils in the
emission process. Or equivalently we can ask what happens if the
surrounding medium reacts very differently according to which mass
eigenstate is emitted. Are there then oscillation phenomena?

This is a question of ``quantum damping'' or
``decoherence''~\cite{me}. It corresponds, in the extreme case, to
conditions in which it would not be possible to form the coherent
initial state Eq.~(\ref{threeB}).  It results from the fact that in
forming the density matrix for the beam, we are instructed to average
(or ``trace'') over the many unobserved variables of the source or
equivalently the ``recoil detectors.''  Now if the conditions are such
that the unobserved variables go into different states according to
which neutrino mass eigenstate is emitted, the result will be a
strongly mixed or incoherent state for the beam and not a pure state
like Eq.~(\ref{threeB}).  In the extreme case of strong damping and so
no coherence between mass eigenstates 1 and 2, the density matrix in
the mass eigenstate basis, for an emitted $\nu_e$ would have zero
off-diagonal elements, reflecting no coherence between mass
eigenstates. The diagonal elements would have the values
$\cos^2\theta$ and $\sin^2\theta$ in terms of the usual mixing
angles~\cite{raffelt}, reflecting the amount of $\nu_e$ in the two
mass eigenstates. The state breaks up into an incoherent mixture of
mass eigenstates, since the mass has been ``measured''~\cite{me}, and
there will be no oscillations.

For the very small mass differences usually contemplated in mixing
processes, however, the resulting differences in momentum are so small
relative to the spread of momentum in the surroundings (detailed
calculations are possible by the methods of Ref.~\cite{me}) that the
resulting decoherence will be negligible. Note analogous results will
follow if some property other than mass is ``measured'' by the
surroundings, in which case there would be oscillations, in general.

\section{General Remarks}

With these points in mind we can now deal with some of the issues
raised in the introduction.

{\it Lines, Continua, and ``Length of the Wavepacket''.}---From the
present point of view---always assuming stationarity---a ``line'' is
simply a strong source in a narrow energy range, and no particular
coherence properties should be assigned to it. The ``length of the
wavepacket,'' if we wish to use that language, is simply determined by
the energy (strictly, momentum for non-relativstic particles) band
used in the data sample. For a ``line'' we are, aside from fine
points, essentially interested in the width, regardless of how it
originates.

In principle, then, one can achieve the same results as for a ``line''
with a broad source and energy selection by the detector, if the
resolution of the detector and the number of events is adequate.  For
example, solar neutrino detectors with some degree of energy
resolution could possibly look for oscillations in other regions of
the spectrum in the same way as in the $^7$Be proposal (see~below).

{\it Distinguishability of Beams.}---We stress that in our usual
understanding of quantum mechanics there is no way to even ascribe a
difference to two simple stationary beams with the same energy
spectrum, since they have one and the same density matrix. Hence any
experiment, even one involving detectors with time dependent
elements~\cite{golub}, which could establish such a difference would
be of the utmost interest. This is of course not limited to neutron
physics. An experiment which could tell, say, if an unpolarized photon
beam were made of a mixture of circular and not linear polarizations
would also be very~surprising.

\section{Neutrino Oscillations}

Turning now to searches for neutrino oscillations, the stationarity
assumption must certainly be good for the sun or reactors.  Even a
supernova, evolving relatively rapidly, can be taken as a sequence of
approximately thermal, quasistationary states. Thus we calculate all
effects by an incoherent average over energy.

{\it Coherence and energy resolution.}---Since we need only take an
incoherent average over the energy spectrum, the only relevant
question for the observability of neutrino oscillations is the energy
spread $\Delta E$ in a sample. That is, $\Delta E$ must be
sufficiently small so that the energy dependent phase difference
$\phi_1-\phi_2$ for mass eigenstates $1$ and~$2$ does not vary by more
than $2\pi$ over the sample. At a distance $d$ from the source and in
terms of the oscillation length parameter $l$ given by the inverse of
the momentum difference for relativistic particles $1/l=\Delta
M^2/2E$, one has $\phi_1-\phi_2= d/l$. This leads to the requirement
\begin{equation}\label{two}
\Delta E /E < 2\pi l/d\,.
\end{equation}
(Taking $l$ much larger than the dimensions of the source, otherwise
there is a further average over $d$.)

{\it Separation of Packets.}---Viewed classically, neutrino states
with different masses will move apart because of their differing
velocities, and in Ref.~\cite{reinartz} the fact that for supernovas
the very long flight paths can lead to a macroscopic separation of
wavepackets was analyzed. In the case of a macroscopic separation
there will obviously be no interference effects between mass
eigenstates. On the other hand if the classical separation is very
small, as for most terrestial experiments, one presumably need not
discuss this problem.  Still, one might wonder how small is small
enough and in the case of small or partial separation, if some
additional treatment is perhaps necessary.

None however is in fact needed, at least in the stationary case, since
all effects are taken care of by the average over the energy spectrum.
Since the ``length of the wavepacket'' is simply given by the band of
energies in the sample, the question of the ``separation'' can be
viewed as simply another form of the resolution condition
Eq.~(\ref{two}): Using $v_1-v_2\approx 1/lE$ the classical separation
$s$ is $d/lE$ so Eq.~(\ref{two}) can be written as
\begin{equation}\label{three}
\Delta E < 2\pi/s
\end{equation}
which is the usual result in quantum mechanics that there is no
interference if the two mass packets separate by more than the inverse
of the available resolution. This also indicates, as usual, that
``separation of the wavepackets'' can be compensated by an increase in
the experimental resolution.  Were M\"ossbauer effect-like detection
and resolution ever possible for neutrinos, then very small $l$ could
be studied for various sources. Note that even a very great energy
resolution does not necessarily imply the ability to distinguish mass
eigenstates, unless the associated momentum difference can be
manifested in some way (see the discussion on ``decoherence").

{\it The $^7$Be line.}---Here $\Delta E$ is determined by the spread
of the line, which a rough thermal estimate would give as about 1 KeV.
A detailed calculation by Bahcall~\cite{bahc} roughly verifies this,
but results in a more complicated and asymmetric line shape.

In ``length of the wavepacket'' language the ``length'' here is then
about $1/{\rm KeV}$ or $ 2\times10^{-8}~{\rm cm}$ (natural units), not
too far from the ``length of the wavepacket'' estimate of
$6\times10^{-8}~{\rm cm}$ given in Ref.~\cite{nuss}, (but somewhat
different from that of Ref.~\cite{wilz}). Hence the practical
consequences should be about the same as in Ref.~\cite{nuss}, but for
detailed analysis one can use the exact line shape.  We stress that it
would be wrong to consider ``coherent'' effects like the natural line
width and ``incoherent'' effects like Doppler broadening on a
different footing, only the full energy spectrum is of interest.

The main question here is how small a candidate $l$ can be before the
condition Eq.~(\ref{two}) is violated.  With the figure of about 1~KeV
we have $\Delta E /E\approx 1~{\rm KeV}/1~{\rm MeV}=10^{-3}$, so
Eq.~(\ref{two}) is easily satisfied for $l/d\approx .035$, which is
the seasonal variation in the earth's orbit. To study the possibilites
for smaller $l$ with precision ~\cite{pak}, one can integrate over the
lineshape of Ref.~\cite{bahc}.

{\it Resolution by the Detector.}---As mentioned above, we might also
consider looking at other regions of the solar spectrum using the
resolution of the detector. This requires at least $\Delta E /E\approx
2\pi (.035)\approx .2$, perhaps not totally impracticable, depending
on the detector.

{\it Oscillations for supernova neutrinos.}---Here $l$ must be very
big to be observable, since $d$ is so large, and to see an effect by
moving the detector it must be moved a distance comparable to $l$.
Thus a search for oscillations by moving the detector, or rather with
detectors in different locations, does not come into question.
However, since we now see that the ``length of the wavepacket'' poses
no problem, and that the only condition is Eq.~(\ref{two}), we can
return to the proposal of Ref.~\cite{reinartz}, where one uses the
energy resolution of the detector to look for oscillations as function
of energy. Given some degree of energy resolution, one divides the
events from a supernova burst into energy bins, and is sensitive to
oscillation lengths according to Eq.~(\ref{two}).  Certainly the
condition $l\sim d$ seems very improbable for an supernova; on the
other hand it is amusing that there is a way, in principle, to see
such tiny mass differences.

\section{Quantum Optics}

Our main point has been that for a stationary density matrix all
information is in the energy spectrum. This is generally true---up to
the question of non-trivial degeneracy---but it should be borne in
mind that in quantum optics, there is another dimension to the energy
spectrum.  For particle beams like neutrons and neutrinos we usually
ignore the possibility of effects related to degeneracy or fermi
statistics---certainly a very good approximation---so that the energy
spectrum simply refers to the distribution of single particle
energies. The ``energy spectrum'' and the ``color spectrum'' are the
same. On the other hand in quantum optics (where a state of three red
photons can have the same energy as a state of one blue photon) there
is also an energy spectrum for a given color or mode of the field,
namely the distribution of photon number in the given
mode~\cite{sigl}.

The role of the single particle density matrix is played by the first
order correlation function~\cite{glauber}, $G^1$ which determines the
intensity or single counting rate. Here again it is possible to see
that if this quantity is constant in time it is determined simply by
the intensity spectrum, that is by the average number of particles in
each mode, paralleling the situation for single particle quantum
mechanics.

When we consider the possibilty of multiply occupied states and
counting correlation measurements however, there is a new aspect in
that this single counting quantity may be stationary although the
state as a whole is not. Consider a single mode in the Glauber or
coherent state. The average number of photons or the single counting
rate is constant. But the {\it field} has a time dependence, so in
principle there is a measurement (e.g.~of the electric field) showing
a time dependence. Here there are significant phase relations between
states with different occupation numbers; there is more information
than simply the occupation numbers. (Non-stationarity should not
really be surprising here since at low frequencies, that is in the
maser, the coherent state is in fact used as a clock.)  On the other
hand the ``chaotic state'' of quantum optics, which resembles the
thermal state, is completely characterized by its occupation numbers,
and any observable is stationary, not just the single counting rate.

This point is significant in qualifying the question as to the
distinguishability of two beams with the same energy spectrum.  If we
allow for the study of correlations between counts, then, as for
example in the ``bunching'' of photons in quantum optics, it might be
said that ``packets'' are observable, and thus that two beams with the
same intensity spectrum could nevertheless be distinguished by such
correlations. However these ``bunches'' are not the wavepackets of
single particle quantum mechanics, applicable to low density beams,
for which the discussion of distinguishability originally was
intended.


\section*{Acknowledgments}

I am indebted to H.J.~Lipkin for several discussions and help in
clarifying the discussion of mixing.  I would also like to thank
G.~Raffelt for many discussions and for stressing that the energy
spectrum ought to be enough, as well as R.~Glauber for a similar
remark, and finally R.~Golub for a discussion concerning neutron
physics.


\end{document}